# Super-resolving phase measurement with short wavelength NOON states by quantum frequency up-conversion

Zhi-Yuan Zhou,[1,2,†,*] Shi-Long Liu,[1,2,†] Shi-Kai Liu,[1,2] Yin-Hai Li,[3] Dong-Sheng Ding,[1,2] Guang-Can Guo,[1,2] and Bao-Sen Shi[1,2,*]

[1]CAS Key Laboratory of Quantum Information, USTC, Hefei, Anhui 230026, China
[2]Synergetic Innovation Center of Quantum Information & Quantum Physics, University of Science and Technology of China, Hefei, Anhui 230026, China
[3]Department of Optics and Optical Engineering, University of Science and Technology of China, Hefei, Anhui 230026, China

*Corresponding author: zyzhouphy@ustc.edu.cn; drshi@ustc.edu.cn

[†]These two authors contributed equally to this work.

Precise measurements are the key to advances in all fields of science. Quantum entanglement shows higher sensitivity than achievable by classical methods. Most physical quantities including position, displacement, distance, angle, and optical path length can be obtained by optical phase measurements. Reducing the photon wavelength of the interferometry can further enhance the optical path length sensitivity and imaging resolution. By quantum frequency up-conversion, we realized a short-wavelength two-photon number entangled state. Nearly perfect Hong–Ou–Mandel interference is achieved after both 1547-nm photons are up-converted to 525 nm. Optical phase measurement of two-photon entanglement state yields a visibility greater than the threshold to surpass the standard quantum limit. These results offer new ways for high precision quantum metrology using short wavelength quantum entanglement number state.

Important in nearly all fields of science, precision measurements have been the long pursued goal in physical science and a window to discoveries. Quantum metrology takes advantages of quantum entanglements to advance precision measurements beyond that achievable with classical methods [1]. A particularly useful class of states are maximally path-entangled multi-photon states (NOON states) $|N::0\rangle = 1/\sqrt{2}(|N0\rangle_{AB} + |0N\rangle_{AB})$, which contains $N$ indistinguishable particles in an equal superposition with all particles being in one of the paths $A$ and $B$. By measuring the phase with a $N$-particle NOON state, the precision of the measurement is $\Delta\phi = 1/N$ --the Heisenberg limit; for measurements with $N$ uncorrelated photons, the precision of the measurement is $\Delta\phi = 1/\sqrt{N}$ -- the standard quantum limit (SQL). This super-sensitivity of the phase measurement has many important applications including imaging, microscopy, gravity-wave detection, measurement of material properties, and chemical and biological sensing [2-4].

The $N\phi$ dependence of the $N$-particle NOON state is a manifestation of the $N$-photon de Broglie wavelength $\lambda/N$ [5]. Referred to as phase super-resolution, this dependence is responsible for an interference oscillation $N$ times faster than that of a single photon. The observation of the reduced de Broglie wavelength has been reported for two to six photons [6–15]. Observing the reduced de Broglie wavelength does not guarantee beating SQL [11, 13]. The maximum photon number surpassing SQL reported is four [13], most of the previous experiments that beat SQL are limited to two photons. To reduce the effective de Broglie wavelength, another effective way is to reduce the fundamental wavelength of the photon. Short wavelength photons can further enhance the sensitivity in optical path length measurement and reduce diffraction effects in imaging. Quantum frequency conversion (QFC) is a promising technique to change the frequency of a photon while retaining its quantum properties [16]. Much progress has been achieved for QFC covering aspects such as performing complete Bell-state measurement for polarization state in quantum teleportation [17], up-conversion of the telecom-band time-bin entanglement state [18], single photon generation from quantum dot [19], squeezing vacuum states [20], orbital angular momentum single photon and entangled states [21, 22], fabrication of up-conversion single-photon detectors [23], and development of frequency domain Hong–Ou–Mandel (HOM) interference [24]. To date, the shortest wavelength entanglement generated is 390 nm polarization entanglement, which is based on biexciton resonant hyper-parametric scattering in a semiconductor [25]. By using QFC technique, the achieved wavelength of the entanglement state can be even shorter than 390 nm. Most of the previous works have mainly focused on frequency conversion of a single part of the photon pair, to frequency conversion of both photons and sequentially processing the converted photons has not been studied before.

In this work, using the QFC technique, both photons in the pair at 1547-nm are up-converted to 525-nm. The indistinguishability of the photon pair after up-conversion is preserved and demonstrated using nearly perfect HOM interference between the two 525-nm photons. Phase measurements of the 525-nm two-photon NOON state generated yield a visibility greater than the threshold to surpass the SQL. Such 525-nm two-photon NOON states are equivalent to a 1547-nm six-photon NOON states. In addition, direct up-conversion of the 1547-nm two-photon Fock-state to 525-nm is also demonstrated. These results suggest new techniques for high precision measurements based on short-wavelength NOON states of high photon number.

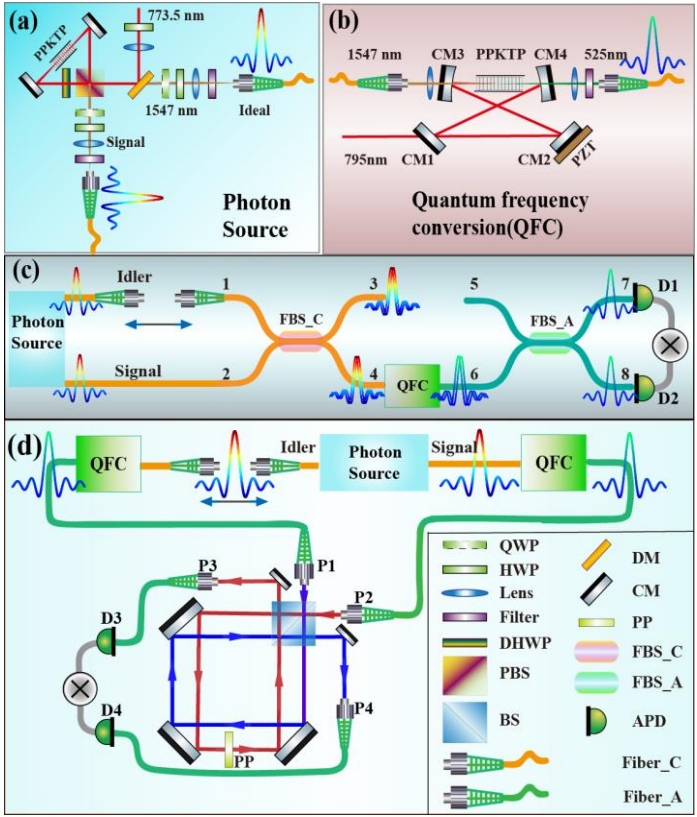

FIG. 1. Experimental setup for each experiment. (a) Simplified diagram of the single-photon source used in experiments; (b) Key module used for frequency up-conversion, in which a 1547-nm photon is focused into the center of a 2-cm type-I PPKTP crystal inside a ring cavity of single resonance at 795-nm, and is up-converted to a 525-nm photon that is collected by a single-mode fiber for detection or further processing. (c) Diagram for QFC of two-photon Fock state. (d) Setup for up-conversion of both 1547-nm photons to generate the 525-nm two-photon NOON state and perform phase measurements of this state in a self-stable tilted Sagnac interferometer. M: mirrors; QWP (HWP): quarter (half) wave plate; DHWP: double half wave plate at 773.5 nm and 1547 nm; FC: fiber coupler; PP: phase plate; BS: beam splitter; CM: cavity mirror; DM: dichromatic mirror; PBS: polarized beam splitter; APD: avalanched single-photon detector.

QFC can be accomplished using sum frequency generation (SFG), in which the annihilation of a strong pump photon ($\omega_p$) and a weak signal photon ($\omega_1$) creating a SFG photon with frequency ($\omega_2 = \omega_1 + \omega_p$). The effective Hamilton operator for this process is [16]

$$\hat{H}_{eff} = i\hbar\xi(\hat{a}_1\hat{a}_2^\dagger - \hat{a}_1^\dagger\hat{a}_2), \qquad (1)$$

where $\hat{a}_1$ and $\hat{a}_2^\dagger$ represent, respectively, the annihilation and creation operators of the signal and SFG photons; $\xi$ is a constant, which is proportional to the product of the pump amplitude $E_p$ and the second-order susceptibility $\chi^{(2)}$. The evolution of $\hat{a}_j$ obtained in the Heisenberg's picture is given as:

$$\hat{a}_1(t) = \hat{a}_1(0)\cos(\xi t) - \hat{a}_2(0)\sin(\xi t), \qquad (2)$$

$$\hat{a}_2(t) = \hat{a}_2(0)\cos(\xi t) + \hat{a}_1(0)\sin(\xi t). \qquad (3)$$

When $\xi t_f = \pi/2$, the input signal field is completely converted to the output SFG field $\hat{a}_2(t_f) = \hat{a}_1(0)$. As $\xi$ strongly depends on the pump amplitude, the key point for reaching maximum conversion efficiency is to increase the pump power. In this letter, the conversion efficiency is increased using a ring cavity to enhance the pump power.

We first up-convert the two-photon Fock state $|02\rangle_{1547nm}$ generated from the HOM interference of the 1547-nm telecom-band photon pair. In the experimental setups (Fig. 1), the 1547-nm photon source [Fig. 1(a)] used is generated from a 2-cm length type-II periodically poled potassium titanyl phosphate (PPKTP) crystal in a Sagnac-loop interferometer, which is pumped using a 90-mW 773.5-nm laser. The source is of high brightness and compact and high entanglement quality. The detail performance of the source can be found in Ref. [26], the difference from Ref. [26] is that the crystal length is increased from 1-cm to 2-cm, which has a narrower photon emission bandwidth of 1.28-nm. In these experiments, the pump laser of the source is from a self-building doubling laser, which can provide more than 300-mW power of 773.5-nm laser beam. The source emitted degenerate photon pairs at 1547-nm. By rotating the wave plate at the input port of the Sagnac-loop, the pump power is circulating in the clockwise direction, therefore polarization orthogonal photon pair is generated, the pair is separated by PBS and collected to single mode fibers (SMF). The polarization of the photon inside the SMFs is controlled by two groups of wave plates at the output ports behind the PBS.

The dependence of the internal quantum efficiency on circulation power for the QFC module is characterized by using a coherent continuous narrow bandwidth laser source as the signal. The results are shown in Fig. 2. The maximum quantum efficiency measured is 0.37 for pump power of 660-mW. In the sequent experiments, the internal conversion efficiencies for both cavities are maintained at around 0.16 because the total output power of our Ti: sapphire laser is limited to 1.5 W.

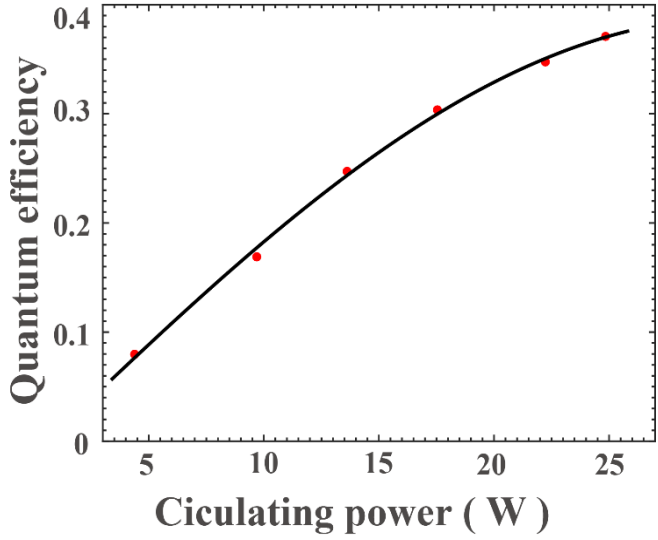

FIG. 2. Internal quantum conversion efficiency as a function of circulation power inside the cavity.

To generate the two-photon Fock state for up-conversion, the signal and idler photons from the photon source are first brought to interfere at a telecom-band fiber beam splitter (FBS_C). Next, the output ports 3 and 4 of the FBS_C are connected to two avalanche photon detectors (APD, ID220, free running InGaAs detector, 20% quantum efficiency, 5-μs dead time, 3-kcps dark counts), from which the output signals are sent to a coincidence device (Timeharp 260, 1.6-ns coincidence window) for coincidence measurement. From the pattern of the HOM interference [Fig. 3(a)] (coincidences between port 3 and 4), the visibilities is (97.90±0.25)% and the two-photon coherent length is 0.83-mm, which corresponds to a bandwidth of 1.28-nm. The high HOM visibility of the source guarantees that the output ports 3 and 4 of FBS_C are in the maximally entangled state of $1/\sqrt{2}(|20\rangle+|02\rangle)_{1547nm}$. Next, another telecom-band FBS is connected to port 3. The post selection state of this second FBS is $|11\rangle_{1547nm}$, the output of which is connected to the APDs for coincidence measurement; the results are given in Fig. 3(b). When the two photons completely overlap at the first FBS, coincidences between the second FBS are double than that when the two-photon are completely non-overlapping. This phenomenon indicates that the output of the first FBS is indeed a two-photon Fock state at the dip of the HOM interference.

Then the output photon from port 4 is injected into the QFC module [Fig. 1(b)] for frequency up-conversion. The QFC module consists of a 2-cm type-I sum-frequency-generation (SFG) PPKTP crystal inside a ring cavity with a single resonance at 795-nm. The 1547-nm photon is focused into the center of the crystal, and the up-converted 525-nm photon is collected in a single-mode fiber for detection or further processing. The pump powers are fixed at 550 mW and 520 mW for the two cavities in the experiments to balance the conversion efficiency of the QFCs. The single photon at 1547 nm only single passes the cavity mirrors CM3 and CM4. The ring cavity has two flat mirrors (CM1, CM2) and two concave mirrors (CM3, CM4) with curvature of 80-mm. The input mirror CM1 has 3% transmittance for 795 nm; CM2 is high reflective-coated at 795 nm (R>99.9%); mirror CM3 is high transmitted-coated at 1547-nm (T>98%) and high reflective-coated at 795-nm; mirror CM4 is high-reflective-coated at 795 nm, high transmitted-coated at 1547-nm and 525-nm (T>%98). The beam waists are about 40 μm at the center of the PPKTP crystals. PPKTP crystals have dimensions of 1-mm×2-mm×20-mm. The phase matching temperatures of the two crystals are kept at 51°C and 49.5°C, respectively. The pump beams are modulated with two electro-optical modulators for locking the cavity using Pound-Drever-Hall method (PDH) [27].

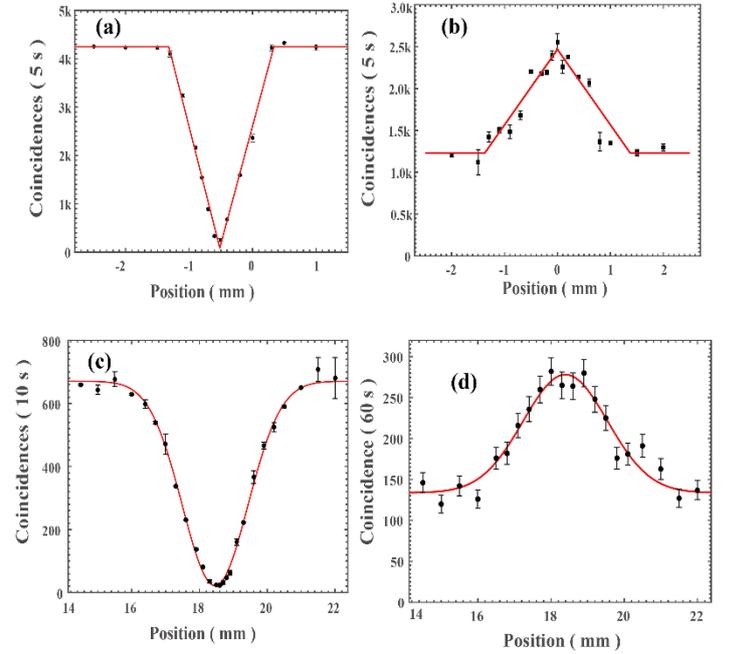

FIG. 3. Experimental results for two-photon Fock state up-conversion. (a) and (c) HOM interference patterns of the source and after up-conversion. (b) and (d) Coincidence counts after the second FBS for the source and for up-converted photons. Error bars in (d) are evaluated assuming the photon detection process obeys Poisson statistics.

Two kinds of measurements are performed on the up-converted 525-nm photon. Coincidence counts between the 1547-nm photon (port 3) and the 525-nm photon (port 6) are measured as a function of the position of the air gap [Fig. 3(c)]. Nearly perfect HOM interference is observed with visibility of (96.72±0.82)%, the fitted curve has a Gaussian shape reflecting the filtering effect in the SFG process [28], for qualitative discussions about the change of the shape of the HOM interference curve, please refer to supplementary information for detail. The SFG bandwidth of the crystal is narrower than the bandwidth of the photon pair (0.5 nm for the 1547-nm photon), which can be seen from the broadening of the bandwidth of the HOM curve. As for the photon source, the up-converted 525-nm photon is split using a 525-nm FBS; the output signals of the FBS are sent to two visible Si APDs (50% quantum efficiency, 300 cps dark count) for coincidence measurements. The results [Fig. 3(d)] show the same behavior as in Fig. 3(b) and demonstrate that the quantum properties of the photon after QFC have been preserved.

Next, both 1547-nm photons are sent to two QFC modules [Fig. 1(d)] for up-converting to 525-nm photons. To demonstrate the indistinguishability of the up-converted 525-nm photons, HOM interference is performed first for the two 525-nm photons where the output signals of the two QFCs are connected to a 525-nm FBS (not shown in Fig. 1(d)). The interference pattern [Fig. 4(a)] is almost perfect with visibility (97.66±0.91)%, indicating that the up-converted photons are highly indistinguishable. High HOM-interference visibility guarantees that a maximal two-photon NOON state of the form

$1/\sqrt{2}(|20\rangle+|02\rangle)_{525nm}$ is generated.

Finally, the up-converted two 525-nm photons are injected into a tilted self-stable Sagnac-interferometer for one-photon and two-photon interference measurements [13]. The self-stable interferometer is the key for long-term data acquisition, which is rather different from traditional Mach-Zehnder that needs active stabilization if the measurement time is very long. The interferometer contains a phase plate (PP) to vary the phase in one optical path. For one-photon interference, one input port of the interferometer is blocked. A highly attenuated 1547-nm laser beam of single-photon level is used to generate the 525-nm attenuated single photons. The interference fringes obtained as a function of rotation angle of the phase plate [Fig. 4(b)] give visibility (97.51±0.5)%. For the two-photon inputs from both input ports of the interferometer, the two-photon interference fringes [Fig. 4(c)] have an oscillation period that is half that of the single photon interference; the visibility of the two-photon interference is (84.93±3.18)%, which surpasses the SQL threshold (71%) for two photon NOON states.

Now we will estimate the overall detection efficiency of the photon pairs after being generated from the PPKTP crystal. The average collection efficiency of signal and idler photons of the entangled source is 0.24. In the experiments, the efficiencies include: filter transmission (0.80); total transmission of the faces of the crystal, cavity mirror (CM3, CM4), lenses and wave plates is 0.86; the fiber collection efficiencies of the 525-nm SMF is about 0.60; the bandwidth of the SFG is 0.5 nm, which reduced the total quantum efficiency to 0.064; The single photon detection efficiency is 0.50; the transmission of the air gap is 0.8; the in and out transmission of the self-stable Sagnac-interferometer is about 0.51. Therefore, the overall detection efficiency of the signal photon is about $2.0\times10^{-6}$.

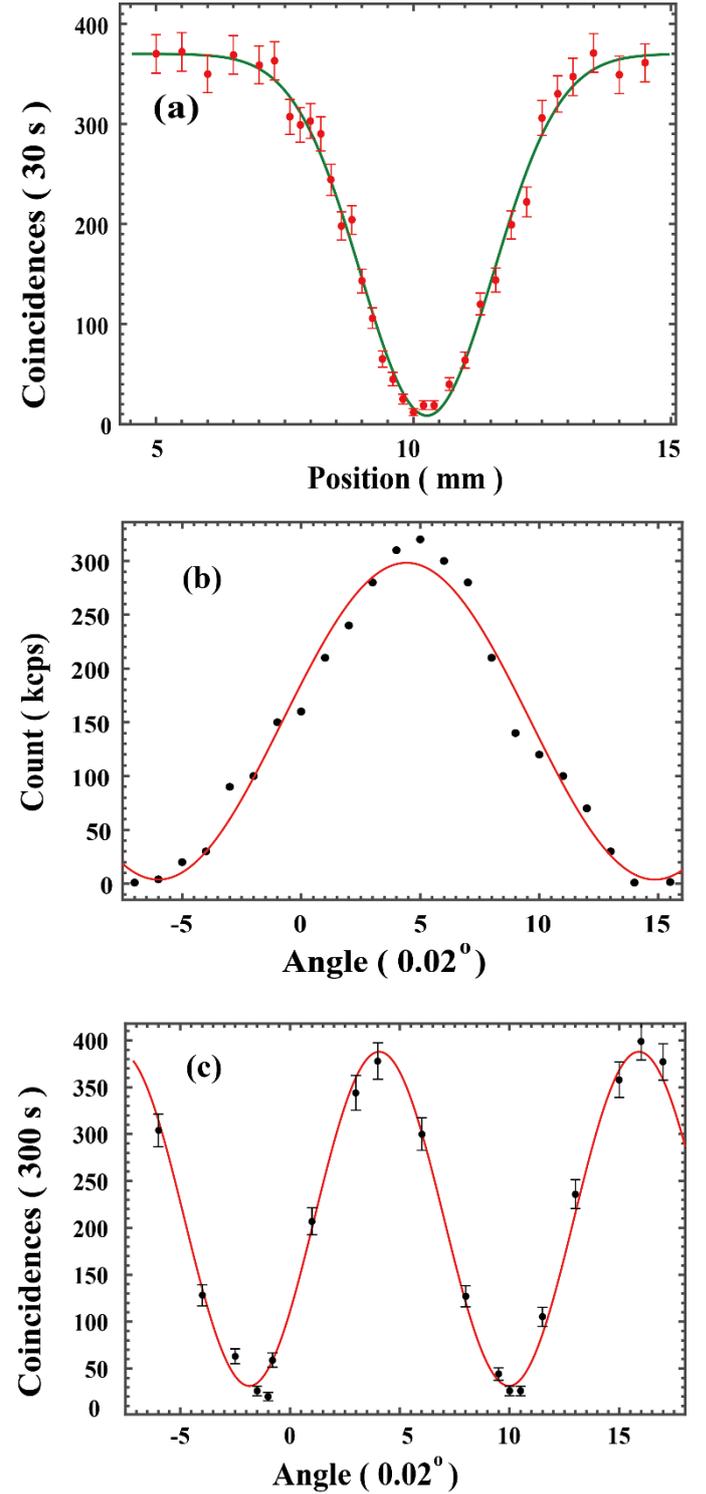

FIG.4. Experimental results for the generation of 525-nm two-photon NOON states. (a) HOM interference for up-converted 525-nm photon pair. (b) and (c) one-photon and two-photon interference patterns as function of rotation angle of the phase plate. Error bars in (a) and (c) are evaluated assuming the photon detection process obeys Poisson statistics.

For summary, QFC for the two-photon Fock state and both photons in the pair are realized for the first time in this work. Our results demonstrate that the single-photon properties and the indistinguishability are preserved during frequency conversion. Using QFC, we demonstrated a promising technique to generate a short-wavelength multi-photon NOON state that has a similar effect in increasing the photon number at the fundamental photon wavelength. The de Broglie wavelength of the 525-nm two-photon NOON state thus generated is equivalent to a 6-photon NOON state at 1547-nm. The 525-nm two-photon interference has visibility high enough to surpass the SQL.

The remaining problem to be solved for high photon number conversion is the quantum conversion efficiency. If this efficiency reaches unity, higher-photon-number NOON states at longer wavelengths can be directly converted to short-wavelength NOON states. The efficiency of the conversion can be solved using longer SFG crystals and a better cavity mirror coating with lower cavity losses. Our method can be easily extended to high-photon-number NOON states if the conversion efficiency is increased. We want to point out that by changing the crystal parameters and the pump wavelength, the up-converted photon wavelength can be extended to UV regime, which can be shorter than 390 nm photon pairs reported in Ref. [25]. Once the phase plate used in the experiment is replaced by transparent optical materials, we can determine their optical properties such as dispersion, absorption and surface imaging with high precision, which may beat the SQL. This work paves the way for QFC of high-number multi-particle entangled states, offering the prospect of new techniques for high-precision quantum metrology.

This work is supported by National Natural Science Foundation of China (Grant Nos. 11604322, 61275115, 61435011, 61525504, and 61605194), China Postdoctoral Science Foundation (Grant No. 2016M590570) and the Fundamental Research Funds for the Central Universities.

# Supplementary information

**Changes of the shape of the HOM interference in QFC**

For spontaneous parametric down conversion (SPDC) in a type-II PPKTP crystal, the phase mismatching $\Delta k_{II}$ can be expressed as:

$$\Delta k_{II} = k_p - k_s - k_i + \frac{2\pi}{\Lambda_{II}} \quad (1)$$

Where $k_j = 2\pi n_j / \lambda_j (j = p, s, i)$ is wave vector inside the crystal, $\Lambda_{II}$ is the periodically poling period of the crystal. For a narrow bandwidth continuous pump laser beam, the degenerate emission spectra of the crystal can be expressed as

$$F(\lambda_s, \lambda_i, \lambda_p) \propto Sinc^2(\frac{\Delta k_{II} L}{2}) \quad (2)$$

Where $L$ is the crystal length. By inserting the Sellmeier equation for $n_y$ and $n_z$ into equation (2), [S1, S2] we can obtain the emission spectra for the generated photon pairs. The numerical simulation results is given in figure S1(a), the bandwidth of the numerical calculation is about 1.3 nm. Because the spectra of the signal and idler photon are $Sinc^2$ function, therefore the HOM interference pattern should be a reversed triangle, as the HOM interference shape is the Fourier transformation of the joint spectra of the signal and idler, one can refer to Ref. [28] for detail.

The SFG crystal used in our experiments is a type-I quasi-phase matching PPKTP crystal, the phase mismatching can be expressed as

$$\Delta k_I = k_{SFG} - k_p - k_s + \frac{2\pi}{\Lambda_I} \quad (3)$$

Where $k_j = 2\pi n_j / \lambda_j (j = SFG, p, s)$ is wave vector inside the crystal. $\Lambda_I$ is the periodically poling period of the crystal. For a narrow bandwidth continuous pump laser beam, the acceptance spectra of the signal photon can be expressed as

$$G(\lambda_s, \lambda_i, \lambda_p) \propto Sinc^2(\frac{\Delta k_I L}{2}) \quad (4)$$

Where $L$ is the crystal length. By inserting the Sellmeier equation fo $n_z$ into equation (2), [S2] we can obtain the emission spectra for the generated photon pairs. The numerical simulation results is given in figure S1(b), the bandwidth of the numerical calculation is about 0.5 nm.

The SFG process can be viewed as a spectra filtering process for the signal and idler photon, therefore effective spectra for the photon pairs after SFG can be expressed as

$$\Gamma(\lambda_s,\lambda_i,\lambda_p) \propto F(\lambda_s,\lambda_i,\lambda_p)G(\lambda_s,\lambda_i,\lambda_p)^2$$
$$= Sinc^2(\frac{\Delta k_{II} L}{2})Sinc^4(\frac{\Delta k_I L}{2}) \quad (5)$$

The simulation of the spectra of the up-converted photon pair is shown in figure s1(c). After being up-converted, the spectra of the photon pair will change from *Sinc²* function to quasi-Gaussian function, and it will be well approximated by using Gaussian function. This is the reason to use Gaussian function to fit the HOM interference data for the up-converted photon pair.

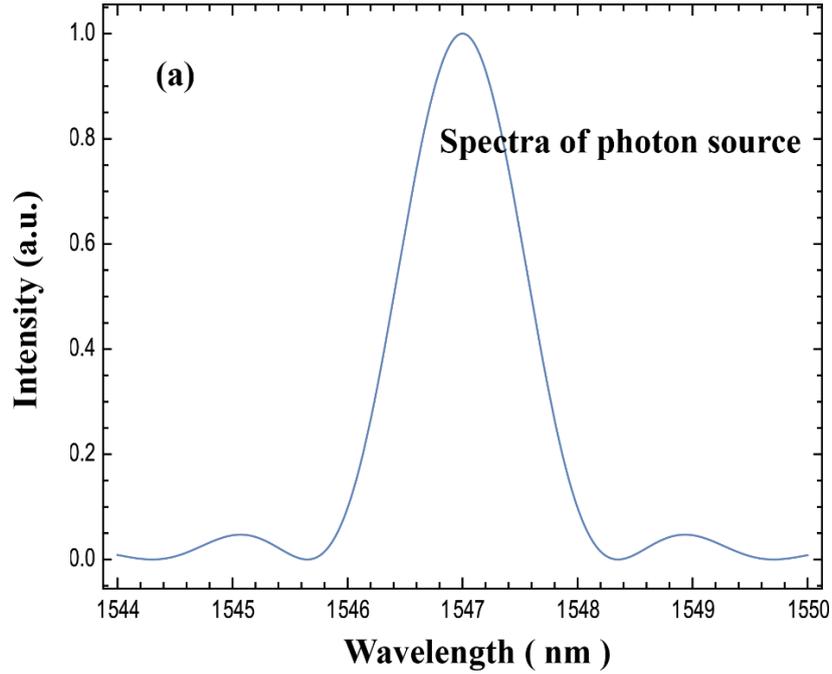

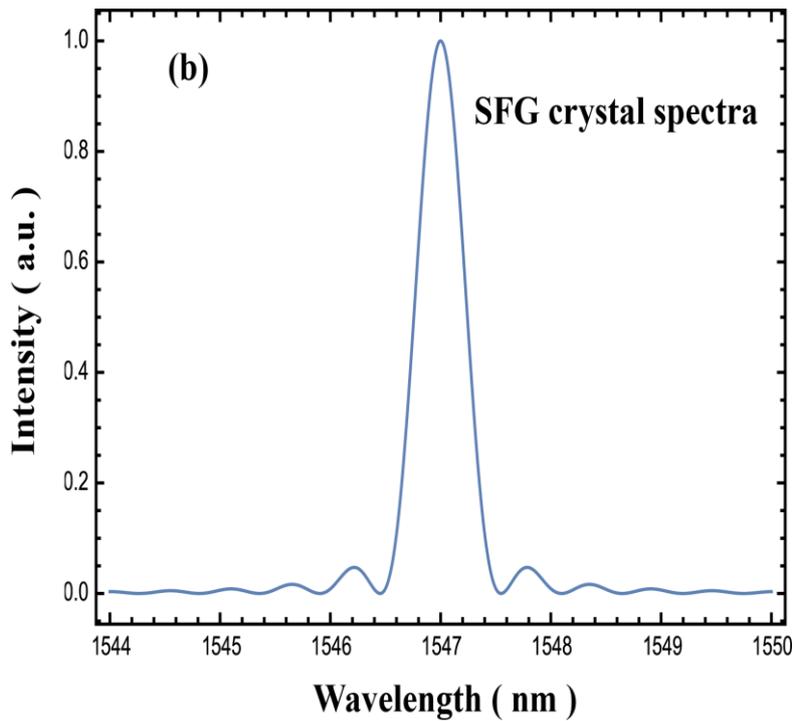

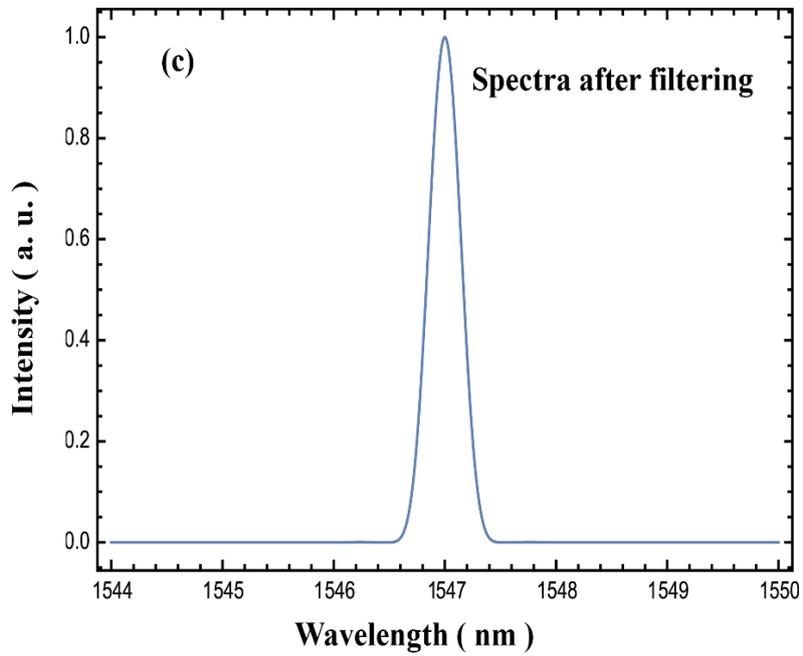

Fig. S1. (a) The emission spectra of the signal and idler photon generated from type-II PPKTP crystal based on SPDC. (b) The acceptance spectrum for the SFG crystal. (c) The filtered spectra for the photon pair after being up-converted.